\documentclass[a4paper,11pt]{article}
\pdfoutput=1

\usepackage{amssymb,amsmath,bm}
\usepackage{a4wide}
\usepackage{color}
\usepackage{slashed}
\usepackage{graphicx}
\usepackage[latin1]{inputenc}
\usepackage{amsfonts}
\usepackage{lscape}
\usepackage{amsthm}
\usepackage{booktabs}
\usepackage{array}
\usepackage{rotating}
\usepackage[numbers, sort]{natbib}
\usepackage{multirow}
\usepackage{float}

\usepackage[small,bf]{caption}
\newcommand{\TeV}{{\rm TeV}}
\newcommand{\eps}{\epsilon}
\newcommand{\ord}[1]{\mathcal{O}\left( #1 \right)}
\newcommand{\gm}{\gamma^\mu}

\newcommand{\sbL}{\overline{s}_L}
\newcommand{\bbL}{\overline{b}_L}

\newcommand{\cbL}{\overline{c}_L}

\newcommand{\sbR}{\overline{s}_R}
\newcommand{\bbR}{\overline{b}_R}

\newcommand{\cbR}{\overline{c}_R}

\newcommand{\dbX}{\overline{d}_X}
\newcommand{\sbX}{\overline{s}_X}
\newcommand{\bbX}{\overline{b}_X}

\newcommand{\dbY}{\overline{d}_Y}
\newcommand{\sbY}{\overline{s}_Y}
\newcommand{\bbY}{\overline{b}_Y}

\newcommand{\mubL}{\overline{\mu}_L}
\newcommand{\mubR}{\overline{\mu}_R}

\newcommand{\mubX}{\overline{\mu}_X}

\newcommand{\ebL}{\overline{e}_L}
\newcommand{\ebR}{\overline{e}_R}

\newcommand{\ebX}{\overline{e}_X}
\newcommand{\ebY}{\overline{e}_Y}

\newcommand{\Lo}{L}

\begin{document}

\vspace{1cm}
\begin{titlepage}
\vspace*{-1.0truecm}
\begin{flushright}
MPP-2012-70\\
CERN-PH-TH/2012-085\\
TUM-HEP-832/12
\end{flushright}

\vspace{0.8truecm}

\begin{center}
\boldmath

{\Large\textbf{Universal Constraints on Low-Energy Flavour Models}}

\unboldmath
\end{center}

\vspace{0.4truecm}

\begin{center}
{\bf Lorenzo Calibbi$^a$, Zygmunt Lalak$^{b,c}$,\\
Stefan Pokorski$^{b,d}$, Robert Ziegler$^{d,e}$
}
\vspace{0.4truecm}

{\footnotesize

$^a${\sl Max-Planck-Institut f\"ur Physik (Werner-Heisenberg-Institut),
 F\"ohringer Ring 6, \\ D-80805 M\"unchen, Germany}\vspace{0.2truecm}

$^b${\sl Institute of Theoretical Physics, Faculty of Physics, University of Warsaw, Ho\.za 69,\\ 00-681, Warsaw, Poland}\vspace{0.2truecm}

$^c${\sl CERN Physics Department, Theory Division,\\ CH-1211 Geneva 23, Switzerland \vspace{0.2truecm}}

$^d${\sl TUM-IAS, Technische Universit\"at M\"unchen,  Lichtenbergstr.~2A,\\ D-85748 Garching, Germany \vspace{0.2truecm}}

$^e${\sl Physik Department, Technische Universit\"at M\"unchen,
James-Franck-Stra{\ss}e, \\D-85748 Garching, Germany}

}
\end{center}

\vspace{0.4cm}
\begin{abstract}
\noindent
It is pointed out that in a general class of flavour models one can identify certain universally present FCNC operators, induced by the exchange of heavy flavour messengers. Their coefficients depend on the rotation angles that connect flavour and fermion mass basis. 
The lower bounds on the messenger scale are derived using updated experimental constraints on the FCNC operators. 
The obtained bounds are different for different operators and in addition they depend on the chosen set of rotations.
Given the sensitivity expected in the forthcoming experiments, the present analysis suggests interesting room  for discovering new physics. As the highlights emerge the leptonic processes,  $\mu\rightarrow e\gamma$, $\mu\rightarrow eee$ and $\mu\rightarrow e$ conversion in nuclei.
\end{abstract}

\end{titlepage}


\section{Introduction}

One of the reasons to go beyond the Standard Model (SM) is to explain the observed hierarchies in fermion masses and mixing. Any physics beyond the SM is expected to contain new sources of flavour violation and is therefore strongly constrained by experimental data.  An interesting hypothesis, known  under the name of Minimal
Flavour Violation(MFV), is that in the extensions of the SM , as in the SM itself, the maximal fermion flavour symmetry
$SU(3)^5$ is broken only by the Yukawa couplings~\cite{MFV}.  Under that hypothesis, the new sources of flavour violation can be  described  in terms of higher dimension effective operators, symmetric under $SU(3)^5$, with Yukawa couplings included as spurion fields. The smallness of some of the Yukawa couplings provides a strong suppression of the FCNC and CP violating effects and therefore MFV is consistent  with a new  scale as low as a few TeV.  However, in most explicit models aiming at explaining the Yukawa coupling pattern constructed so far, the initial flavour symmetry group is much smaller than $SU(3)^5$.  One may then expect potentially more dangerous sources of  the FCNC effects and, in consequence, the need for a much higher UV completion scale. It is the purpose of this paper to investigate that question   in a large class of flavour models based on horizontal symmetries and to find experimental observables most sensitive to such completions.

Similarly as under the MFV hypothesis, once a particular flavour symmetry is assumed,  one can construct all effective low-energy operators using  a spurion analysis, with the symmetry breaking fields playing the role of spurions~\cite{LPR}. Such an analysis has to be repeated for each chosen symmetry group.
Furthermore, although in principle all operators allowed by the symmetry arguments are expected to be generated
by the (unspecified) UV completions, their coefficients do depend on the UV dynamics and therefore are not under
control in such an approach.

In this paper we do not take that path. Instead, we point  out that in a general class of flavour models one can identify certain patterns of universally present effective FCNC operators. They depend  only on the rotation angles transforming the light fermion fields from the flavour
basis to the mass eigenstate basis, and the minimal bounds on the scale of UV completions can be estimated for
different qualitative pattern of the rotation angles.  This is possible because in the flavour basis certain operators
are unavoidably present as a consequence of the dynamics supposed to explain  the hierarchical Yukawa couplings. For instance, if some flavour diagonal effective operators with non-equal coefficients were always present, our point would be immediately obvious. As we shall see, it is easy to identify the minimal set of operators in the flavour diagonal basis and the above conclusion remains valid.  Thus, one can  determine the minimal FCNC effects which do not depend on the details of the flavour symmetry.  We investigate models based on abelian and non-abelian flavour symmetries.

Effective higher dimension operators such as 4-fermion operators or 2-fermion penguin operators necessarily originate from integrating out bosonic degrees of freedom. Here we analyse the flavour violating effects arising from an exchange of heavy scalar degrees of freedom (possibly together with heavy fermions) with the mass scale $M$ that are an integral part of many flavour models. These effects are suppressed by $1/M^2$. We obtain an absolute lower bound on the scale $M$ of the order of 20 TeV
in generic abelian flavour models. In non-abelian models, on the other hand, additional suppression factors in the FCNC effects allow $M$ to be as low
as the TeV scale.

There are of course also other scalar bosonic degrees of freedom whose integration out generates flavour violating higher dimensional operator. Always present are the SM vector or Higgs bosons. The four-fermion operators generated at tree-level are  then suppressed by $1/M^2$ and additional powers of $v^2/M^2$ since they arise from SU(2)$_L$ breaking effects. The case in which these effects are the only new source of flavour-violation has been studied in \cite{BGPZ}, implying a bound on the scale of the order of roughly 1 TeV. In supersymmetric models
flavour violation enters low energy physics through sparticle exchange. Since the suppression scale is only $1/m_{\rm SUSY}^2$, these effects can be relevant even if the flavour sector lives at very large scales provided the SUSY breaking scale is even higher \cite{High-Energy}. Finally,  four-fermion operators could  arise from an exchange of  low-energy flavour gauge bosons as it  has been recently studied in \cite{Grinstein, Andrzej}. 

\bigskip 

\noindent In Section 2 we shortly review the UV completion of flavour models in order to show that there must always exist heavy fields that couple to the light fermions
and we study the minimal, unavoidable effective flavour violating operators which arise from the exchange of these messenger fields. In section 3 we compare the obtained universal pattern of the minimal  FCNC effects with  the  experimental bounds on flavour-violating operators and obtain constraints on the messenger scale depending on light rotation angles. We then make the additional assumption of a messenger sector compatible with SU(5), which allows us to include other relevant operators in the discussion leading to a variety of correlations between experimental observables. We finally conclude in section 4. 


\section{Model Setup}

\label{FN-UVC}

We consider models with a general flavour symmetry group\footnote{For our analysis it is irrelevant whether the group is discrete, global or gauged.} $G_F$ spontaneously broken by the vevs of 
scalar fields $\phi_I$ that will be called flavons in the following. The SM Yukawa couplings arise from higher-dimensional $G_F$ 
invariant operators involving the flavons \cite{FN,MMM1,MMM2} (see also \cite{Berezhiani:1990wn}):
\begin{align}
\mathcal{L}_{yuk} & = y_{ij}^U ~\overline{q}_{L i} u_{Rj}\, \tilde{h} + y_{ij}^D~ \overline{q}_{L i} d_{Rj} \, h +{\rm h.c.} & 
 y_{ij}^{U,\,D} & \sim \prod_{I} \left( \frac{\langle\phi_I\rangle}{M} \right)^{n^{U,\,D}_{I,ij}}, 
\label{yuk}
\end{align}
where the suppression scale $M \gtrsim \langle\phi_I\rangle $ is the typical scale of the flavour sector dynamics. The coefficients of the effective operators are assumed to be $\ord{1}$, so that the hierarchy in the Yukawa matrices arise exclusively from the small order parameters $\epsilon_I \equiv {\langle\phi_I\rangle}/{M}$. 
The transformation properties of the SM fields and the flavons under $G_F$ are properly chosen, so that $\epsilon_I$ together with their exponents $n^{U,D}_{I,ij}$ reproduce the observed hierarchy of fermion masses of mixing.

\begin{figure}[t]
\centering
\includegraphics[scale=0.7]{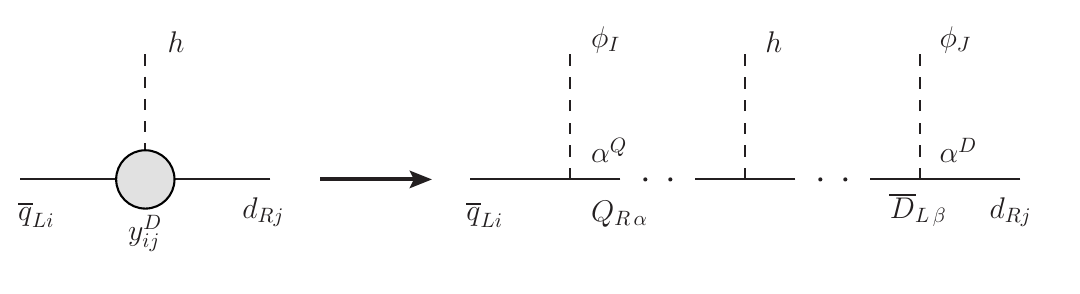}
\caption{Schematic diagram for the Fermion UVC.\label{fig-FUVC}}
\end{figure}

\begin{figure}[t]
\centering
\includegraphics[scale=0.7]{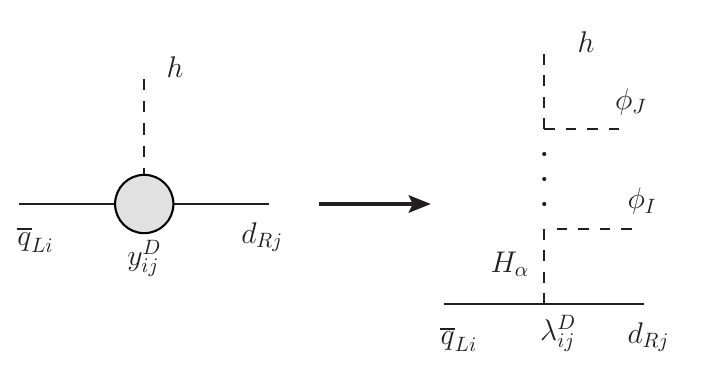}
\caption{Schematic diagram for the Higgs UVC.\label{fig-HUVC} }
\end{figure}

In order to UV-complete these models one has to ``integrate in'' new heavy fields at the scale $M$. These messenger fields are vectorlike and charged under $G_F$. In order
to generate the effective Yukawas of Eq.~(\ref{yuk}), the messengers must couple to SM fermions and flavons and, depending on their nature, 
they mix either with the SM fermions or with the SM Higgs. 
In the first case one has to introduce vectorlike fermions with the quantum numbers of the SM fermions (see Fig.~\ref{fig-FUVC}). 
In the second case one introduces scalar fields with the quantum numbers of the SM Higgs field (see Fig.~\ref{fig-HUVC}). 
In the fundamental theory, small fermion masses arise from a small mixing of light and heavy fermions for the first possibility, while they arise from small vevs of the heavy scalars in the second case. 
We refer to these two possibilities as ``Fermion UV completion'' (FUVC) and ``Higgs UV completion'' (HUVC), respectively.

The only interactions that are relevant for our discussion are the ones that involve messenger fields and SM fermions. The rest of the Lagrangian is only 
responsible for generating the mixing between light and heavy states and do not affect the minimal flavour effects we are going to discuss in the next section.
The relevant interactions can be readily seen in Figs.~\ref{fig-FUVC} and \ref{fig-HUVC}. 
In particular, in the case of FUVC we have couplings among (mainly) light and heavy fermions of the schematic form:
\begin{align}
\label{Lfermion}
\mathcal{L} & \supset \alpha^Q ~   \overline{q}_{L i} {Q_R}_\alpha \,\phi_I + \alpha^D ~\overline{D}_{L \beta} d_{R j} \,\phi_J +{\rm h.c.},       
\end{align}
while in the HUVC we are only interested in interactions involving SM fermions and Higgs messengers:
\begin{align}
 \label{Lhiggs}
 \mathcal{L} & \supset \lambda^D_{ij}~ \overline{q}_{L i} d_{R j} \, H_\alpha  +{\rm h.c.}
\end{align}
Since the hierarchy is supposed to arise from the flavour symmetry breaking alone, we can assume that all dimensionless couplings in the fundamental Lagrangian are $\ord{1}$. A more detailed discussion of the structure of the messenger sector is presented in \cite{High-Energy}.

\subsection{Model-universal FCNC effective operators}

\begin{figure}[t!]
\centering
\includegraphics[scale=0.65]{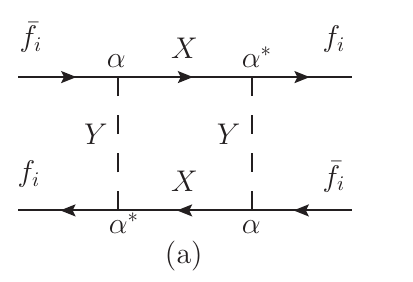}
\includegraphics[scale=0.6]{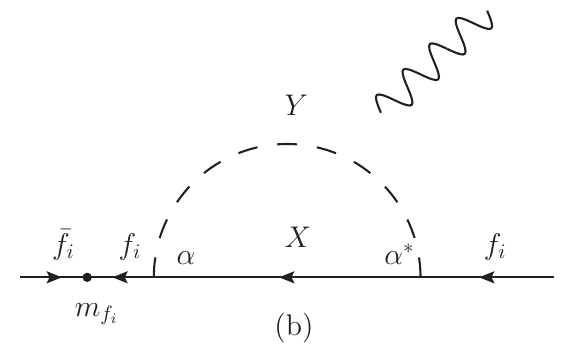}
\includegraphics[scale=0.65]{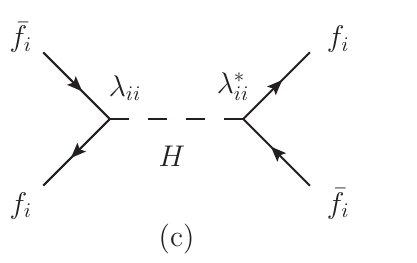}
\caption{Schematic diagrams responsible for the arising of flavour-violating operators\label{fig:diag}.}
\end{figure}

 We now want to derive the effective flavour-violating operators that arise from messenger exchange independently of the details of the particular flavour model. For this we consider flavour-conserving operators, which then induce FCNC effects in the mass basis that only depend on light rotation angles.  

Let us first assume the presence of a coupling in the messenger Lagrangian of the form 
\begin{equation}
{\cal L} \supset \alpha \, \overline{f}_{Li} X_R Y,
\end{equation}
where $\alpha \sim \ord{1}$, $f_{Li}$ is a (mainly) light fermion and $X_R$ and $Y$ are a fermion and a scalar of which at least one is a heavy messenger, cf.~Eqns.~(\ref{Lfermion}, \ref{Lhiggs}). From the box diagram with $X,Y$ propagating in the loop (see Fig.~\ref{fig:diag}a) we get the effective operator 
\begin{equation}
\label{DeltaS2}
{\cal L}_{eff} \supset \frac{|\alpha|^4}{16 \pi^2 M^2}  (\overline{f}_{Li} \gamma^\mu f_{Li})^2,
\end{equation}
where $M$ is the heaviest mass in the loop and we neglected factors of $\ord {1}$. We can use the same coupling also to write down a penguin diagram with a mass insertion in the external fermion line (see Fig.~\ref{fig:diag}b). This generates the dipole operator
\begin{equation}
\label{mueg}
{\cal L}_{eff} \supset \frac{|\alpha|^2}{16 \pi^2 M^2}  m_i \, \overline{f}_{Li} \sigma^{\mu \nu} f_{Ri} F_{\mu \nu},
\end{equation}
where $m_i$ is the light fermion mass and we have again estimated the coefficient very roughly. 

The couplings $ {\cal L} \sim \alpha \, \overline{f}_{Li} X_R Y$ are indeed present in the messenger sector of every flavour model. This is clear at least for $i=1,2$, since the light generations have to couple to some messengers in order to obtain a full rank mass matrix. The same argument holds for $i=3$ in the down\footnote{Except in models where the bottom mass arises at the renormalisable level, like in 2HDM with large $\tan{\beta}$.} and charged lepton sector, but in general not in the up sector, since the top quark can get massive without coupling to the messenger sector. 

Finally there are also certain tree-level operators which unavoidably arise in Higgs UV completions. Consider for example the couplings of the heavy neutral Higgs fields to the down quarks 
\begin{equation}
{\cal L} \supset \lambda_{ij} \overline{d}_{Li} d_{Rj} H.
\end{equation}
In order to get a full rank down mass matrix, at least one of the four couplings $\lambda_{11}, \lambda_{12}, \lambda_{21},\lambda_{22}$ has to be non-zero. By integrating out $H$ at tree-level (see Fig.~\ref{fig:diag}c) we therefore obtain at least one of the four operators 
\begin{equation}
\label{4ops}
{\cal L}_{eff} \supset \frac{|\lambda_{ij}|^2}{M^2} (\overline{d}_{Li} d_{Rj}) (\overline{d}_{Rj} d_{Li}) \qquad (i,j=1,2) 
\end{equation}
Again the same reasoning goes through in the up 1-2 sector, and in every charged lepton and down sector (e.g. at least one of the four operators in Eq.~(\ref{4ops}) must exist also for $i,j=1,3$ and $i,j = 2,3$), but in most generality not in sectors involving the top quark. 

In principle there are tree-level operators also in fermion UV completions arising from flavon exchange. As can be seen from Eq.~(\ref{yuk}), these operators are suppressed by $m_i m_j / M_{\phi}^2 \langle \phi \rangle^2$, where $m_i$ are the light fermion masses,  $M_\phi$ is the flavon mass and $\langle \phi \rangle$ its vev. For flavon mass and vev not far from the messenger scale this contribution scales with the fourth power of the inverse messenger scale and gives only very mild bounds on the messenger scale \cite{MMM1}. We therefore neglect these contributions in the following. 

Having collected the unavoidable flavour-conserving operators in Eqns.~(\ref{DeltaS2})--(\ref{4ops}), we now go to the mass basis using approximate transformations of this kind:
\begin{equation}
d_{Li} \to d_{Li} + \sum_{j \neq i} \theta^{DL}_{ij} d_{Lj}\,,
\label{eq:rot}
\end{equation}
in order to obtain flavour-violating operators. 
In specific models there are in general other contributions to such operators that could be even larger. However, the contribution discussed above does not
depend on the details of the flavour symmetry and its breaking pattern and therefore allows to estimate minimal predictions for the operator coefficients.

Notice that in abelian models there is no reason for cancellations among different contributions to a given FCNC effective operator generated by Eq.~(\ref{eq:rot}), 
since the operators in Eqns.~(\ref{DeltaS2})--(\ref{4ops}) arise from integrating out flavour messengers that by construction have different $\ord{1}$ couplings to light fermions, i.e.~the breaking of flavour universality is O(1).
In non-abelian flavour models those couplings are universal (controlled by the symmetry) and  the breaking of flavour universality is suppressed by small order parameters. This is because in those models the operator of Eq.~(\ref{DeltaS2}) in the flavour basis is given by (restricting to the 1-2 sector)
\begin{align}
{\cal L}_{eff} \sim |\alpha|^4 \left( \overline{f}_{L1} \gamma^\mu f_{L1} +  \overline{f}_{L2} \gamma^\mu f_{L2}  \right)^2,
\end{align}
which clearly remains in this form after rotating to the mass basis. Flavour transitions are only generated upon including a universality breaking term
\begin{align}
{\cal L}_{eff} \sim |\alpha|^4 \left( \overline{f}_{L1} \gamma^\mu f_{L1} +  \overline{f}_{L2} \gamma^\mu f_{L2}  + \Delta_{12}  \overline{f}_{L2} \gamma^\mu f_{L2} \right)^2.
\end{align} 
Each flavour transition in the operator in the mass basis is then suppressed by $\Delta_{12}$ that depends on the flavon vevs that are responsible for universality breaking. Therefore in non-abelian models there is an additional suppression of the above effect that depends on the particular flavour transition. In principle this suppression can be as large as in MFV, but in models with a single non-abelian factor, like in most explicit models constructed so far, the additional suppression factor is rather mild compared to MFV. 
To estimate this factor we consider the case of a SU(3)$_F$ model \cite{King-Ross} with all quarks transforming as a ${\bf 3}$ and flavons as ${\bf \overline{3}}$.\footnote{In models with both singlets and triplets there is no additional suppression of flavour transitions involving singlets, but possibly larger suppression for transitions involving only triplets.}
The flavons get hierarchical vevs that induce the quark masses. The flavon vev responsible for universality breaking in the $i$-$j$ sector is therefore roughly given by the square root of the Yukawa coupling $\sqrt{y_{jj}}$. The situation is similar in the case of some U(2)$_F$ flavour models \cite{u2-models}, where transitions in 1-3 and 2-3 sector are always unsuppressed. 

Taking into account the general case of two Higgs doublet models (2HDM), one finds the following suppression factors for each flavour transitions (two flavon insertions):
\begin{table}[H]
\centering
\begin{tabular}{|c | c | c|}
Flavour transition & Suppression factor in SU(3)$_F$ & Suppression factor in U(2)$_F$\\
\hline
$u - t$ & $\ord{1}$ & $\ord{1}$ \\  
$c - t$ & $\ord{1}$ & $\ord{1}$ \\  
$u - c$ & $\eps^4$  & $\eps^4$ \\  
\hline
$d - b$ & $ \eps^3 \tan{\beta}$ & $\ord{1}$\\  
$s - b$ &  $ \eps^3 \tan{\beta}$ & $\ord{1}$\\  
$d - s$ &  $ \eps^5 \tan{\beta}$ &  $\eps^5 \tan{\beta}$ \\  
\hline
\end{tabular}
\caption{\label{suppNA}Additional suppression factors of flavour transitions in simple non-abelian models.}
\end{table}

\noindent where $ \tan{\beta} = v_u/v_d$ and $\eps$ is of the order of the Cabibbo angle. This table implies for example that the coefficients of the four-fermion operators
relevant for $K-\overline{K}$ mixing
 would get an additional suppression of $\eps^{10} \tan{\beta}^2$ (we need two flavour transitions) with respect to the abelian case. The model-independent bounds on effective operator coefficients in abelian flavour models can be therefore easily extended to simple non-abelian groups. Moreover, the abelian case is relevant for non-abelian models with some SM fermions transforming as singlets of the non-abelian flavour group.

As an example for discrete flavour models (see \cite{AF} and references therein) we briefly discuss the model in Ref.~\cite{Feruglio} based on a $A_4 \times Z_3 \times U(1)$ lepton flavour symmetry. Left-handed leptons transform as triplets under $A_4$, while the three right-handed leptons  are in distinct one-dimensional $A_4$ representations. Flavor-diagonal operators involving only right-handed leptons are effectively unconstrained by the symmetry and the corresponding LFV processes that arise in the mass basis do not receive additional suppression factors. Instead LFV operators involving left-handed leptons experience additional suppression due to necessary presence of flavon vevs that break $A_4$. The  effects discussed here, which arise from rotating flavour diagonal operators, represent only the minimal source of LFV operators. 
Typically non-abelian discrete flavour symmetries lead to certain selection rules for flavour off-diagonal operators, which however still allow for some unsuppressed LFV processes, in the case of the above model those ones 
which satisfy $\Delta L_e \Delta L_\mu \Delta L_\tau = \pm 2$~\cite{Feruglio}.

We are now going to use the procedure outlined above in order to obtain estimates for the minimal coefficients of FCNC operators induced by the messenger sector of generic flavour
models. The most interesting of these operators are those related to $\Delta F=2$ processes and the LFV decays $\mu \to eee$, $\mu \to e \gamma$.
In Table \ref{minOpsPred} we show the estimates of the operator coefficients separately for Higgs and fermion UVC for abelian models 
and add for comparison the MFV prediction. 

A few comments regarding the estimation of the Wilson coefficients are in order:

\begin{itemize}
\item The expressions in Table \ref{minOpsPred} are valid for abelian flavour models, but can be generalised to simple non-abelian groups using the suppression factors provided in Table \ref{suppNA}. 

\item The expressions involving b-quarks is valid in general only if $m_b$ does not arise at the renormalisable level.

\item The estimate on the tree-level contribution to  $\mu \to e e  e$ in Higgs UVCs is rather conservative, since it accounts for the possibility that $\lambda^E_{11}, \lambda^E_{12}, \lambda^E_{21}$ all vanish. Still this minimal effect can be sizable if the rotations are large. 

\end{itemize}
In summary one can obtain minimal predictions of the coefficients of certain flavour-violating effective operators, which do not depend on the details of the flavour model but only on the light fermion mass matrix. These estimates are derived from the coefficients of flavour-diagonal operators (which can be easily obtained in generic flavour models) and the corresponding rotation angles. In non-abelian models one has to take into account the additional suppression discussed above.

\begin{table}[t]
\centering
\setlength{\extrarowheight}{1pt}
\setlength{\tabcolsep}{2pt}
\setlength{\extrarowheight}{1.5pt}
\begin{tabular}{|c||c|c||c|}

Operator &  Higgs UVC & Fermion UVC & MFV \\ 

\hline

$ (\sbL \gm d_L) (\sbL \gm d_L) $ &
$ \Lo \, \theta_{12}^{DL} \theta_{12}^{DL} $ &  
$   \Lo \, \theta_{12}^{DL} \theta_{12}^{DL}$ & 
$\eps^{10}$
  \\ 

$ (\sbR \gm d_R) (\sbR \gm d_R) $ &
$ \Lo \, \theta_{12}^{DR} \theta_{12}^{DR} $ &  
$   \Lo \, \theta_{12}^{DR} \theta_{12}^{DR}$ & 
$\eps^{10} y_d^2 y_s^2$
  \\

$(\sbL d_R) (\sbR d_L) $ & 
$   \theta_{12}^{DL} \theta_{12}^{DR} $ & 
$ \approx 0$  & 
$\eps^{10} y_d  y_s$
 \\ 

\hline

$ (\cbL \gm u_L) (\cbL \gm u_L) $ &
$  \Lo \, \theta_{12}^{UL} \theta_{12}^{UL} $ &  
$  \Lo \, \theta_{12}^{UL} \theta_{12}^{UL}$ & 
$\eps^{10}  $

  \\ 

$ (\cbR \gm u_R) (\cbR \gm u_R) $ &
$  \Lo \, \theta_{12}^{UR} \theta_{12}^{UR} $ &  
$  \Lo \, \theta_{12}^{UR} \theta_{12}^{UR}$ & 
$\eps^{10} y_b^4 y_u^2 y_c^2$

  \\

$(\cbL u_R) (\cbR u_L) $ & 
$   \theta_{12}^{UL} \theta_{12}^{UR} $ & 
$ \approx 0 $ & 
$\eps^{2} y_d^2 y_s^2 y_u y_c$

\\ 

\hline

$ (\bbL \gm d_L) (\bbL \gm d_L) $ &
$ \Lo \, \theta_{13}^{DL} \theta_{13}^{DL} $ &  
$  \Lo \,  \theta_{13}^{DL} \theta_{13}^{DL}$ & 
$\eps^{6}$

  \\ 

$ (\bbR \gm d_R) (\bbR \gm d_R) $ &
$ \Lo \, \theta_{13}^{DR} \theta_{13}^{DR} $ &  
$  \Lo \,  \theta_{13}^{DR} \theta_{13}^{DR}$ & 
$\eps^{6} y_d^2 y_b^2$
  \\ 
  
$(\bbL d_R) (\bbR d_L) $ & 
$    \theta_{13}^{DL} \theta_{13}^{DR}$  & 
$\approx 0 $& 
$\eps^{6} y_d y_b$
  \\ 

\hline

$ (\bbL \gm s_L) (\bbL \gm s_L) $ &
$ \Lo \, \theta_{23}^{DL} \theta_{23}^{DL} $ &  
$   \Lo \, \theta_{23}^{DL} \theta_{23}^{DL}$ & 
$\eps^{4}$
  \\ 

$ (\bbR \gm s_R) (\bbR \gm s_R) $ &
$ \Lo \, \theta_{23}^{DR} \theta_{23}^{DR} $ &  
$   \Lo \, \theta_{23}^{DR} \theta_{23}^{DR}$ & 
$\eps^{4} y_s^2 y_b^2$
  \\ 

$(\bbL s_R) (\bbR s_L) $ & 
$   \theta_{23}^{DL} \theta_{23}^{DR}$ & 
$  \approx 0  $ & 
$\eps^{4} y_s y_b$
  \\ 

\hline

$ \mubX \sigma^{\mu \nu} e_Y F_{\mu \nu}$ &
$ m_\mu \Lo \,  {\rm max} (\theta_{12}^{E L} , \theta_{12}^{E R})$ &
$m_\mu \Lo \,  {\rm max} (\theta_{12}^{E L} , \theta_{12}^{E R}) $  &  -      
\\

\hline

 $ (\mubX \gm e_X) (\ebX \gm e_X) $ &
$\Lo \,  \theta_{12}^{EX} $ &  
$ \Lo \,   \theta_{12}^{EX}$ & -
 \\
 $ (\mubX e_Y) (\ebY e_X) $  & $ \theta_{12}^{EX}  \theta_{12}^{EL} \theta_{12}^{ER}$ &  $\approx 0 $ & - \\

\hline
\hline

\end{tabular} 
\caption{Relevant operators and their minimal Wilson coefficients in units of $1/M^2$ for HUVC, FUVC and MFV. Here $\Lo \simeq 1/16 \pi^2$, $X, Y = L,R$ with $Y \neq X$.}
\label{minOpsPred}
\end{table}

\section{Phenomenological Implications}

The predictions for the minimal flavour-violating effects obtained in the previous section  are to be compared
with experimental bounds for those operators. In Table \ref{minOps} we list the present bounds on the Wilson coefficients of the relevant operators for a suppression scale of 1 TeV.\footnote{For completeness, we include the bounds to left-right vector operators like $(\sbL \gm d_L) (\sbR \gm d_R)$, even though we do not use them in our analysis, 
as in HUVC they give a negligible constraint compared to the scalar operators like $(\sbL d_R) (\sbR d_L)$, 
while in FUVC they do not arise in a model-independent way. } 

\begin{table}[t]
\centering
\setlength{\extrarowheight}{1pt}
\setlength{\tabcolsep}{2pt}
\setlength{\extrarowheight}{1.5pt}
\begin{tabular}{|c|c||c|c|}

Process & Relevant operators & \multicolumn{2}{c|} {Bound on $c/\TeV^2$} \\ 

& &   Re & Im   \\

\hline
\hline

\multirow{3}{3cm}{\centering $\Delta m_K;~\epsilon_K$} & ~ $ (\overline{s}_X \gm d_X) (\overline{s}_X \gm d_X) $ ~ &
  ~~~~~~$9.0\times 10^{-7}$~~~~~~ &  $3.4\times 10^{-9}$
  \\ 
  & $ (\overline{s}_L \gm d_L) (\overline{s}_R \gm d_R) $ &
  $1.9\times 10^{-7}$ & $7.2\times 10^{-10}$
  \\ 
   & $(\sbL d_R) (\sbR d_L) $ & 
  $4.7\times 10^{-9}$ & $1.8\times 10^{-11}$ \\ 

\hline

\multirow{3}{3cm}{\centering $\Delta m_D;~|q/p|_D, A_\Gamma \ $} &
$ (\overline{c}_X \gm u_X) (\overline{c}_X \gm u_X) $ &
  $4.7\times 10^{-7}$ & $~1.3\times 10^{-7}$ ~[$ 3.8 \times 10^{-9} $]~~\\

& $ (\overline{c}_L \gm u_L) (\overline{c}_R \gm u_R) $ &
  $ 7.4 \times 10^{-7}$&  $~2.1 \times 10^{-7}$ ~[$ 5.9 \times 10^{-9} $]~~\\

&
$(\cbL u_R) (\cbR u_L) $ & 
  $4.1\times 10^{-8}$ &  $1.1 \times 10^{-8}$ [$ 3.3 \times 10^{-10} $]\\ 

\hline

\multirow{3}{3cm}{\centering $\Delta m_{B_d};~S_{\psi K_S}$} &
$ (\overline{b}_X \gm d_X) (\overline{b}_X \gm d_X) $ &
  $2.9 \times 10^{-6}$ & $2.6 \times 10^{-6}$
  \\ 
& $ (\overline{b}_L \gm d_L) (\overline{b}_R \gm d_R) $ &
 $ 4.8 \times 10^{-6}$ & $4.3 \times 10^{-6}$ \\
&
$(\bbL d_R) (\bbR d_L) $ & 
  $4.2\times 10^{-7}$ &  $3.8\times 10^{-7}$  \\ 

\hline

\multirow{3}{3cm}{\centering $\Delta m_{B_s};~S_{\psi \phi}$} &
$ (\overline{b}_X \gm s_X) (\overline{b}_X \gm s_X) $ &
  $6.7\times 10^{-5}$ & $ ~5.7 \times 10^{-5} $ ~[$ 4.1 \times 10^{-6} $]~~\\ 
& $ (\overline{b}_L \gm s_L) (\overline{b}_R \gm s_R) $ &
  $ 1.1 \times 10^{-4} $ & $ ~9.4 \times 10^{-5} $ ~[$ 6.7 \times 10^{-6} $]~~\\
&
$(\bbL s_R) (\bbR s_L) $ & 
  $9.7\times 10^{-6}$ & $ ~8.2 \times 10^{-6} $ ~[$ 5.8 \times 10^{-7} $]~~\\ 

\hline
\hline

\multirow{2}{3cm}{\centering $\mu \to e \gamma $} &
\multirow{2}{3cm}{\centering $ \overline{\mu}_X \sigma^{\mu \nu} e_{Y} F_{\mu \nu}$} & \multicolumn{2}{c|} {\multirow{2}{5cm}{\centering $2.9 \times 10^{-10} ~~ [5.9 \times 10^{-11} ]$}} \\

& & \multicolumn{2}{c|}{}   \\

\hline

\multirow{2}{3cm}{\centering $\mu \to e e e$} & $ (\overline{\mu}_X \gm e_X) (\overline{e}_X \gm e_X) $ &\multicolumn{2}{c|} {$2.3\times 10^{-5} ~~ [2.3\times 10^{-7}]$} 
 \\
& $ (\mubX e_Y) (\ebY e_X) $  & \multicolumn{2}{c|} {$6.5 \times 10^{-5}  ~~ [6.5\times 10^{-7}]$} \\

\hline
\hline

\end{tabular} 
\caption{Relevant processes and corresponding operators with bounds on Wilson coefficients. Values in [ ] are for expected future experimental bounds. $X, Y = L,R$ with $Y \neq X$.}\label{minOps}
\end{table}
The bounds on the Wilson coefficients of the $\Delta F=2$ operators have been obtained as in \cite{GinoNir,PerezTASI}, taking into account the QCD running of the operators (from 1 TeV) and imposing as a condition for the meson mass splittings $(\Delta m)^{\rm NP}\leq (\Delta m)^{\rm exp}$, where $(\Delta m)^{\rm NP}$ represent the flavour messenger contributions and $(\Delta m)^{\rm exp}$ the experimental measured values reported in \cite{PDG}.\footnote{The bounds in the $D-\overline{D}$ system are
in good agreement with the values reported in the literature \cite{DDb}, once different conventions in the definition of the operators are taken into account.} 
For the CPV observables, we imposed $\epsilon_K^{NP}\leq 0.6\times \epsilon_K^{\rm exp}$ in the $K-\overline{K}$ system \cite{GinoNir,PerezTASI}, and in the $B_{d,s}$ and $D$ sectors we required that the total prediction (including the SM contribution with the corresponding uncertainty) is within the 2$\sigma$ experimental ranges
$$
\begin{tabular}{ccc}
$ 0.62 \le S_{\psi K_S} \le 0.72 $  \cite{PDG}, & $ - 0.23 \le S_{\psi\phi} \le 0.53 $  \cite{LHCbpsiphi}, 
& $-7.1\,^{\circ} \le \phi_{12}^D \le 15.8\,^{\circ}$ \cite{HFAG}.
\end{tabular}
$$
The bounds on the imaginary parts of the Wilson coefficients have been taken to be the largest possible values consistent with the above ranges combined with the constraints on $\Delta m$. The future bounds on the Wilson coefficients have been obtained using the expected LHCb sensitivities on CPV observables in the $D$ and $B_s$ sectors given in  \cite{LHCbupgrade}. In particular, we consider a future sensitivity on $|q/p|_D$ at the level of $10^{-3}$, of the order of the naive SM prediction, 
even though we cannot exclude that the SM contribution is actually much larger, due to large long distance uncertainties. Of course, the future bounds reported in the table
are valid under the hypothesis that no CPV is observed with an experimental sensitivity at the level mentioned above.
Note that, in this case, the future bounds in the $D-\overline{D}$ system will be almost as strong as in the $K$ sector. The bounds for the LFV processes have been computed taking into account the recent 90\% CL limit ${\rm BR}(\mu\to e\gamma)< 2.4\times 10^{-12}$ obtained by the MEG experiment \cite{MEG}. The future bounds correspond to the expected sensitivities for BR($\mu \to e \gamma$) $\sim 10^{-13}$ \cite{MEG} and BR($\mu \to e e e$) $\sim 10^{-16}$ \cite{PSI}.

Using the information in Tables \ref{minOpsPred} and \ref{minOps} we can easily estimate the bounds on the messenger scale separately for each flavour transition as a function of the rotation angles.\footnote{We neglect running effects above 1 TeV.}  Of course in the SM only the difference of left-handed rotations in the up and down sector are observable. In particular right-handed rotations are not constrained, though in many flavour models they are roughly of the same order as the left-handed ones. Left-handed rotations have to be smaller or equal than the corresponding CKM entries in the absence of cancellations between up and down sector, which is what we assume in the following. Note that also the left-handed rotations in each sector can be complex, since the field redefinitions are already used to absorb the 5 phases of the CKM matrix given by the difference between up and down sector left-handed rotations. 

For a given quark flavour transition we consider six different cases for the rotation angles: the left-handed rotation is either zero or given by the CKM value for this transition, and the right-handed rotation is either maximal\footnote{Maximal rotation means $1/\sqrt{2}$, but we will simply write  ``1'' in the tables. To calculate the bound we take into account all additional factors of $\sqrt{2}$ that arise in this case.}, zero, or equal to the left-handed rotation.
Since there is no effect when both rotations vanish and the predictions are symmetric in the exchange $L \leftrightarrow R$ , only four combinations are relevant. We also distinguish the general case of complex rotation angles from the specific scenario with vanishing phases, in which case only the bounds from CP conserving $\Delta F =2$ observables apply. Finally we distinguish between fermionic and Higgs messengers, the only difference being that in the Higgs UV completion scalar-scalar operators with 1-2 transitions arise at tree-level while they are negligible in the fermionic messenger case. 

The results are shown in Tables \ref{KDtab} and \ref{Bstab}. These numbers give a rough impression of the importance of the minimal effect that we are discussing
(up to unknown $\ord{1}$ coefficients). Indeed for large and complex right-handed rotations in the down 1-2 sector, this effect is sensitive to Higgs messenger scales up to $\ord{10^5}$ TeV, which is roughly the magnitude of the new physics scale one obtains without any suppression \cite{GinoNir} (up to a factor of 1/$\sqrt{\eps}$). Fermionic messengers give weaker constraints, because of the absence of tree-level effects, but nevertheless are sizable in many cases and can test messenger scales up to $\ord{10^3}$ TeV.

\begin{table}[h]
\centering
\begin{tabular}{|c c || c | c |c |c | }
$\theta^{DL}_{12}$ & $\theta^{DR}_{12}$ & HUVC & HUVC$^*$ & FUVC &FUVC$^*$ \\
\hline
\hline
$\eps$ & 0 & $ 19  $& $310 $ & $ 19 $ & $ 310$ \\
$\eps$ & $\eps$ & $ 3,400  $& $ 54,000 $ & $ 19 $ & $ 310$ \\
$\eps$ & $1$ & $4,900 $& $ 80,000 $ & $42 $ & $680 $ \\
0 & $1$ & $ 42  $& $680 $ & $42 $ & $680 $ \\
\hline
\hline
\end{tabular}

\vspace{0.5cm}

\begin{tabular}{|c c || c | c |c |c | }
$\theta^{UL}_{12}$ & $\theta^{UR}_{12}$ & HUVC & HUVC$^*$ & FUVC &FUVC$^*$ \\
\hline
\hline
$\eps$ & 0 & $ 27 $& $ 51 ~ [300] $ & $ 27 $ & $ 51 ~ [300] $ \\
$\eps$ & $\eps$ & $ 1,100 $& $ 2,200 ~ [13,000]$ & $27$ & $51 ~ [300] $ \\
$\eps$ & $1$ & $1,700 $& $ 3,200 ~[19,000] $ & $58 $ & $110 ~ [650] $ \\
0 & $1$ & $ 58    $& $110 ~ [650] $ & $58 $ & $ 110 ~ [650]  $ \\
\hline
\hline
\end{tabular}

\caption{\label{KDtab}Constraints from $K - \overline{K}$ (up) and  $D - \overline{D}$ (down) mixing  on the messenger scale in TeV for Higgs and fermion UV completions with real and complex (*) rotations angles. Values in [ ] correspond to the expected future experimental sensitivities}.
\end{table}

\noindent The most interesting aspect of these tables regards the 1-2 sector. Since the left-handed rotation must be of the order of the Cabibbo angle ($\approx \eps$) either in the up or in the down sector or both, the messenger scale must be larger than the smallest entry in Table \ref{KDtab}. We therefore obtain an overall minimal bound on the messenger scale given by 19 TeV for the case that the rotation angle is real and comes from the down sector. 
Since in non-abelian models there are additional suppression factors (cf.~Table~\ref{suppNA}), the minimal effects alone do not exclude the possibility that the messenger fields of such models could be as light as a TeV and therefore in the reach of LHC. 

\begin{table}[h]
\centering
\begin{tabular}{|c c || c | c |c |c | }
$\theta^{DL}_{13}$ & $\theta^{DR}_{13}$ & HUVC & HUVC$^*$ & FUVC &FUVC$^*$ \\
\hline
\hline
$\eps^3$ & 0 & $ - $& $ - $ & $  - $ & $  - $ \\
$\eps^3$ & $\eps^3$ & $ 19 $& $ 20 $ & $ - $ & $ - $ \\
$\eps^3$ & $1$ & $ 120 $& $ 130 $ & $ 23 $ & $ 25 $ \\
0 & $1$ & $ 23 $& $25 $ & $ 23 $ & $  25 $ \\
\hline
\hline
\end{tabular}

\vspace{0.5cm}

\begin{tabular}{|c c || c | c |c |c | }
$\theta^{DL}_{23}$ & $\theta^{DR}_{23}$ & HUVC & HUVC$^*$ & FUVC  & FUVC$^*$ \\
\hline
\hline
$\eps^2$ & 0 & $ - $& $ - $ & $ - $ & $ - $  \\
$\eps^2$ & $\eps^2$ & $ 17 $  & $ 18 ~ [69] $ & $ - $ & $ - $\\
$\eps^2$ & $1$ & $52 $& $ 57 ~[210] $  & $ 5 $ & $ 5 ~ [20] $\\
0 & $1$ & $ 5    $& $ 5 ~ [20] $ & $ 5 $ & $ 5 ~ [20] $\\
\hline
\hline
\end{tabular}
\caption{\label{Bstab}Constraints from $B_d - \overline{B}_d$ (up) and $B_s - \overline{B}_s$ (down) mixing on the messenger scale in TeV for Higgs and fermion UV completions with real and complex (*) rotations angles. Values in [ ] correspond to the expected future experimental sensitivities.}
\end{table}

\noindent Despite the fact that the constraints in the 1-3 and 2-3 sector are much weaker, they are still relevant since minimal effects in the 1-2 sector do not imply small effects in other sectors where rotations can be large, giving bounds on the messenger scale comparable or even larger than 19 TeV (cf.~Table~\ref{Bstab}). In particular this is true in non-abelian models where transitions in the 1-2 sector are much stronger suppressed compared to sectors involving the 3$^{\rm rd}$ family,  whereas, in abelian models, it is clear from  comparing Tables \ref{KDtab} and \ref{Bstab} that large effects in the 2-3 sector are only possible if the phases in the 1-2 sector are sufficiently suppressed.
The possibility of large CP violation in $B_s$ mixing is especially interesting, since the experimental sensitivity will be improved by LHCb in the near future. 

For lepton flavour transitions we consider six possibilities for the rotation angles: the right-handed rotation can be maximal, zero or $\eps$, as it can be the case in SU(5), and the left-handed rotation is either maximal, zero, or equal to the right-handed rotation. Since there is no effect when both rotation vanish and the predictions are symmetric in the exchange $L \leftrightarrow R$ , only five combinations are relevant. CP violating effects do not play a role here, but we still distinguish between fermionic and Higgs messengers, again the only difference being that in the Higgs UVC certain operators can arise at tree-level. 

\begin{table}[h]
\centering
\begin{tabular}{|c c || c | c  | }
$\theta^{ER}_{12}$ & $\theta^{EL}_{12}$ & HUVC = HUVC$^*$ & FUVC  = FUVC$^*$ \\
&  &  $\mu \to e \gamma \hspace{1cm} \mu \to 3 e $ & $\mu \to e \gamma  \hspace{1cm} \mu \to 3 e $ \\
\hline
\hline
$\eps$ & 0 & 23 [51] \hspace{1cm}  11 [110] & 23 [51] \hspace{1cm} 11 [110] \\
$\eps$ & $\eps$ & 23 [51] \hspace{1cm}  14 [140] & 23 [51] \hspace{1cm} 11 [110] \\
$\eps$ & $1$ & 34 [75] \hspace{1cm}    42 [420] & 34 [75] \hspace{1cm} 12 [120] \\
$0$ & $1$ & 34 [75] \hspace{1cm} 12 [120] & 34 [75] \hspace{1cm} 12 [120] \\
$1$ & $1$ & 34 [75] \hspace{1cm} 88 [880]  & 34 [75] \hspace{1cm} 12 [120] \\

\hline
\hline
\end{tabular}
\caption{\label{Ltab} Constraints from $\mu \to e \gamma$ and $\mu \to eee $ on the messenger scale in TeV for Higgs and fermion UV completions. Values in [ ] are for expected future experimental bounds on these processes.}
\end{table}

\noindent The results are shown in Table \ref{Ltab}. For large rotation angles the minimal LFV effect is sensitive to messenger scales of the order of 50 TeV with present exclusion limits and up to hundreds of TeV with future limits. 
The most interesting aspect is that the branching ratio ${\rm BR}(\mu \to eee )$ can be substantially enhanced with respect to ${\rm BR}(\mu \to e \gamma )$, provided that the corresponding tree-level operator is sizable. In general, there is a lower bound on the ratio ${\rm BR}(\mu \to eee )$/${\rm BR}(\mu \to e \gamma)$ from the dipole transition $\mu \to e \gamma^*$ approximately given by $6 \times 10^{-3}$  \cite{BurasParide}. 
This bound is saturated in many SM extensions like SUSY, and  is indeed close to the ratio one obtains here when both operators arise at loop-level and have the same angle dependence, as it is the case for fermionic messengers. In the case of Higgs UVC instead $\mu \to eee$ can arise through a tree-level exchange of a heavy Higgs messenger. The minimal effect is suppressed by the product of three rotations, but can win over the loop induced one when rotations are large. In this case  ${\rm BR}(\mu \to eee )$/${\rm BR}(\mu \to e \gamma)$ can be as large as 19 if the rotation angles are maximal. This can be read from Table \ref{Ltab} using the formula
\begin{equation}
\frac{{\rm BR}(\mu \to eee )}{{\rm BR}(\mu \to e \gamma)} = \left( \frac{M_{\mu \to eee}}{M_{\mu \to e \gamma}} \right)^4 \,  \frac{{\rm BR^{exp}}(\mu \to eee )}{{\rm BR^{exp}}(\mu \to e \gamma)},
\label{superF}
\end{equation}
where BR$^{\rm exp}(\mu \to X)$ denotes the present experimental limit and $M_{\mu \to X}$ the corresponding bound on the mass scale of the effective operator shown in the table. There are good prospects to test this ratio in the future. 
In fact, if $\mu \to e \gamma$ is observed at the MEG experiment with ${\rm BR}(\mu \to e \gamma)\gtrsim 10^{-13}$, $\mu \to eee$ generated at tree-level (HUVC) can still have a rate close to the present experimental limit (${\rm BR}(\mu \to eee )< 10^{-12}$), whereas in the case when $\mu \to eee$  arises at loop level, the rate is expected to be ${\rm BR}(\mu \to eee ) \gtrsim 8 \times 10^{-16}$ (irrespectively of the mixing angles), which is still in the reach of the future experimental sensitivity.

As an illustration of the magnitude of the effects we estimate the branching ratios for a reference scale of 19 TeV (as needed to satisfy the quark sector constraints):
\begin{align}
{\rm BR}(\mu \to e \gamma ) & \simeq 5.3 \times 10^{-12} \left( \frac{19 ~\TeV}{M} \right)^4 \left( \frac{{\rm max}(\theta^{EL}_{12}, \theta^{ER}_{12})}{\eps} \right)^2, \\
\nonumber \\
{\rm BR}(\mu \to eee) & \simeq 2.9 \times 10^{-13}  \left( \frac{19 ~\TeV}{M} \right)^4  \left( \frac{{\rm max}(\theta^{EL}_{12}, \theta^{ER}_{12})}{\eps} \right)^2 \left( \frac{\theta^{EL}_{12} }{\eps} \right)^2 \left( \frac{\theta^{ER}_{12} }{\eps} \right)^2,
\end{align}
where we assumed HUVC for the $\mu \to eee$ rate. Notice that the bounds from the quark sector still allow for rates of LFV observables in the reach of running or future experiments. 
\bigskip 

\noindent Let us summarise the main points of the above discussion  (valid up to $\ord{1}$ coefficients):
\begin{itemize}
\item The messenger scale in abelian models has to be larger than 19 TeV. 

\item This minimal bound does not prevent large effects in $B_q - \overline{B}_q$ mixing and LFV decays, because the rotations in the corresponding sectors could be large.

\item In non-abelian models the minimal effects do not exclude messengers at the TeV scale and thus in the reach of LHC.

\item ${\rm BR}(\mu \to eee )$/${\rm BR}(\mu \to e \gamma)$   can be as large as $\ord{10}$ for large leptonic rotations. 
\end{itemize} 


\subsection{Predictions in SU(5)}
\begin{table}[t]
\centering
\setlength{\extrarowheight}{1.5pt}
\begin{tabular}{|l|c||c|}
Process & Relevant operators & Bound on $c/\TeV^2$ \\ 

\hline
\hline

\multirow{2}{3.2cm}{${\rm CR}(\mu \to e$ in Ti)} & $ (\mubX \gm e_X) (\dbX \gm d_X) $ & 
  $  5.7 \times 10^{-6} \, [1.9 \times 10^{-8}]$ 
\\ 

&
$(\mubX e_Y) (\dbY d_X) $ & 
 $  1.8 \times 10^{-6} \, [6.3 \times 10^{-9}]$ 
\\   

\hline

\multirow{2}{3.2cm}{${\rm BR}(K_L \to \mu^+ \mu^-)$} &
$ (\mubX \gm \mu_X) (\sbX \gm d_X) $ & 
  $ 2.6 \times 10^{-4}$ \\

&
$(\mubX \mu_Y) (\sbY d_X) $ & 
 $2.1 \times 10^{-5}$
\\   

\hline

\multirow{2}{3cm}{${\rm BR}(K_L \to e^+ e^-)$} &
$ (\ebX \gm e_X) (\sbX \gm d_X) $ & 
  $1.9 \times 10^{-3} $ \\

&
$(\ebX e_Y) (\sbY d_X) $& 
 $6.9 \times 10^{-7}$ \\   

\hline

\multirow{2}{3cm}{${\rm BR}(K_L \to \mu^+ e^-)$} &
$ (\mubX \gm e_X) (\sbX \gm d_X) $ & 
  $9.8 \times 10^{-6}$ \\

&
$(\mubX e_Y) (\sbY d_X) $ & 
 $5.5 \times 10^{-7}$ \\   

&
$(\mubX e_Y) (\dbY s_X) $ & 
 $5.5 \times 10^{-7}$\\   

\hline

\multirow{2}{3cm}{${\rm BR}(B_d \to \mu^+ \mu^-)$} &
$ (\mubX \gm \mu_X) (\bbX \gm d_X) $ & 
  $4.4 \times 10^{-3} \, [1.4 \times 10^{-3}] $ \\

&
$(\mubX \mu_Y) (\bbY d_X) $ & 
 $1.0 \times 10^{-4} \, [3.2 \times 10^{-5}]$ \\

\hline

\multirow{2}{3cm}{${\rm BR}(B_s \to \mu^+ \mu^-)$} &
$ (\mubX \gm \mu_X) (\bbX \gm s_X) $ & 
  $7.1 \times 10^{-3} \, [6.1 \times 10^{-3}] $ 
\\

&
$(\mubX \mu_Y) (\bbY s_X) $& 
 $1.6 \times 10^{-4} \, [1.4 \times 10^{-4}]$ \\   
  
\hline
\hline

\end{tabular} 
\caption{\label{SU5opsbounds}Relevant processes and corresponding operators with bounds on Wilson coefficients. Values in [ ] are for expected future experimental bounds. $X, Y = L,R; Y \neq X$.}
\end{table}

The number of completely model-independent operators that are subject to sizable constraints is restricted to $\Delta F=2 $ and dipole operators involving 1-2 flavour transitions. Only with additional assumptions one can make further statements on e.g. two-quark-two-lepton ($2q2\ell$) operators. A particular well-motivated and predictive assumption is that the flavour sector is compatible with an (approximate) SU(5) GUT structure, which connects lepton and quark operators and correlates the charged lepton and down quark mass matrix. In particular this implies: (i) the existence of heavy states that couple both to quarks and leptons, so that diagrams as in Fig.~\ref{fig:diag} unambiguously induce ($2q2\ell$) operators; (ii)  $M_d \approx M_e^T$ and therefore $\theta^{DR}_{ij} \approx \theta^{EL}_{ij}$ and $\theta^{DL}_{ij} \approx \theta^{ER}_{ij}$.\footnote{Since we are doing order of magnitude estimates, we neglect the high-energy corrections to this relation, such as the Georgi-Jarlskog factor \cite{GJ}, necessary to correctly account for the low-energy mass ratios of the first two generations leptons and down quarks.}

We list the interesting operators along with the bounds on their Wilson coefficients in Table \ref{SU5opsbounds}. These bounds have been obtained using the formulae in \cite{Davidson}. For the operators contributing to $B_{d,s} \to \mu^+ \mu^-$ decays, we have used the new LHCb 95\% CL limits \cite{LHCb}:
\begin{equation}
{\rm BR}(B_{d} \to \mu^+ \mu^-)< 1.0 \times 10^{-9} \qquad {\rm BR}(B_{s} \to \mu^+ \mu^-)< 4.5 \times 10^{-9}.
\end{equation}
As future bounds on these processes we have taken the values corresponding to the SM predictions, while for $\mu\to e$ conversion in nuclei we have considered the future sensitivity ${\rm CR}(\mu \to e ~ {\rm in~ Ti}) \sim	 5\times 10^{-17}$ \cite{mec}. 

Again we compare the bounds with the predictions based on the minimal effects from messenger exchange. In Table \ref{SU5opsPred} we show the  estimates of the operator coefficients separately for Higgs and fermion UVC and add for comparison the MFV prediction.  
\begin{table}[t]
\centering
\setlength{\extrarowheight}{1.5pt}
\begin{tabular}{|c||c|c||c|}
Operator & Higgs UVC & Fermion UVC & MFV\\ 

\hline
\hline

 $ (\mubX \gm e_X) (\dbX \gm d_X) $ &      
$ {\Lo} \theta_{12}^{EX} $ &  
$   {\Lo}  \theta_{12}^{EX}$ &
-
\\

$(\mubX e_Y) (\dbY d_X) $ & 
$\theta_{12}^{E X} \theta_{12}^{D L} \theta_{12}^{D R} $ &  
$ \sim 0 $ &
-
\\   

\hline

$ (\mubL \gm \mu_L) (\sbL \gm d_L) $ & 
$ {\Lo} \theta_{12}^{DL} $ &  
$   {\Lo} \theta_{12}^{DL}$ &
$\eps^5$
\\ 
	
$ (\mubR \gm \mu_R) (\sbR \gm d_R) $ & 
$ {\Lo} \theta_{12}^{DR} $ &  
$   {\Lo}  \theta_{12}^{DR}$ &
$\eps^5 y_d y_s$
\\ 

$(\mubL \mu_R) (\sbR d_L) $ & 
$ {\rm max} (\theta_{12}^{D L}, \theta_{12}^{D R})$ &  
$ \sim 0 $ &
$\eps^5 y_s$
\\   

$(\mubR \mu_L) (\sbL d_R) $ & 
$ {\rm max} (\theta_{12}^{D L}, \theta_{12}^{D R})$ &  
$ \sim 0 $ &
$\eps^5 y_d$
\\

\hline

$ (\ebL \gm e_L) (\sbL \gm d_L) $ & 
$ {\Lo}\theta_{12}^{DL} $ &  
$   {\Lo}  \theta_{12}^{DL}$ &
$\eps^5 $
\\ 

$ (\ebR \gm e_R) (\sbR \gm d_R) $ & 
$ {\Lo} \theta_{12}^{DR} $ &  
$   {\Lo}  \theta_{12}^{DR}$ &
$\eps^5 y_d y_s$
\\

$(\ebL e_R) (\sbR d_L) $& 
$ \theta_{12}^{DL} \theta_{12}^{E L} \theta_{12}^{E R}$ &  
$ \sim 0 $ &
$\eps^5 y_s$
\\   

$(\ebR e_L) (\sbL d_R) $& 
$\theta_{12}^{DR} \theta_{12}^{E L} \theta_{12}^{E R}$ &  
$ \sim 0 $ &
$\eps^5 y_d $
\\
\hline

$ (\mubX \gm e_X) (\sbX \gm d_X) $ & 
$ {\Lo} \theta_{12}^{EX}  \theta_{12}^{DX}$ &  
$   {\Lo}  \theta_{12}^{EX}  \theta_{12}^{DX}$  &
-

\\

$(\mubX e_Y) (\sbY d_X) $ & 
$ \theta_{12}^{E Y} \theta_{12}^{D X}$ &  
$ \sim 0 $ &
-

\\

$(\mubX e_Y) (\dbY s_X) $ & 
$ \theta_{12}^{E Y} \theta_{12}^{D Y}$ &  
$ \sim 0 $ &
-

\\   

\hline

$ (\mubL \gm \mu_L) (\bbL \gm d_L) $ & 
$ {\Lo}\theta_{13}^{DL} $ &  
$  {\Lo}  \theta_{13}^{DL}$ &
$ \eps^6 $
\\ 
	
$ (\mubR \gm \mu_R) (\bbR \gm d_R) $ & 
$ {\Lo} \theta_{13}^{DR} $ &  
$  {\Lo}  \theta_{13}^{DR}$ &
$ \eps^6 y_d y_b$
\\ 

$(\mubL \mu_R) (\bbR d_L) $ & 
${\rm max}(\theta_{12}^{D L}\theta_{23}^{D R},\theta_{12}^{D R}\theta_{23}^{D L}) $ &  
$ \sim 0 $ &
$ \eps^6 y_b$
\\   

$(\mubR \mu_L) (\bbL d_R) $ & 
$ {\rm max}(\theta_{12}^{D L}\theta_{23}^{D R},\theta_{12}^{D R}\theta_{23}^{D L}) $ &  
$ \sim 0 $&
$ \eps^6  y_d$
\\

\hline

$ (\mubL \gm \mu_L) (\bbL \gm s_L) $ & 
$ {\Lo} \theta_{23}^{DL} $ &  
$   {\Lo}  \theta_{23}^{DL}$ &
$ \eps^4 $
\\ 
	
$ (\mubR \gm \mu_R) (\bbR \gm s_R) $ & 
$ {\Lo} \theta_{23}^{DR} $ &  
$   {\Lo}  \theta_{23}^{DR}$ &
$ \eps^6 y_s y_b$
\\

$(\mubL \mu_R) (\bbR s_L) $& 
$ {\rm max}(\theta_{23}^{D L},\theta_{23}^{D R}) $ &  
$ \sim 0$ &
$ \eps^4 y_b$
\\   

$(\mubR \mu_L) (\bbL s_R) $& 
$ {\rm max}(\theta_{23}^{D L},\theta_{23}^{D R}) $ &  
$ \sim 0$ &
$ \eps^4 y_s $
\\   
  
\hline
\hline

\end{tabular} 
\caption{\label{SU5opsPred}Relevant operators and their minimal Wilson coefficients in units of $1/M^2$ for HUVC, FUVC and MFV. Here $\Lo \simeq 1/16 \pi^2$, $X, Y = L,R; Y \neq X$. }
\end{table}
For each flavour transition we then use this table to calculate the bounds on the messenger scale for given rotation angles and look for correlations. Now only four combinations of rotation angles are relevant and again we distinguish between fermionic and Higgs messengers, the only difference being that in the Higgs UVC some operators can arise at tree-level. 

\begin{table}[h]
\centering

\begin{tabular}{|c c ||c c | c | c | c| c|}
$\theta^{DL}_{12}$ & $\theta^{DR}_{12}$ & $\Delta m_K$ & $\eps_K $ & $K_L \to \mu^{\pm} e^{\mp}$ & $\mu \to e \gamma$ & $\mu \to eee$ & $ \mu \to e $ in Ti\\
\hline\hline
$\eps$ & 0 & 19 & 310 & 5.8 & 23 [51] & 11 [110]& 16 [280]\\
$\eps$ & $\eps$ & 19 & 310 & 5.8 & 23 [51] &  11 [110] & 16 [280]\\
$\eps$ & $1$ & 42 & 680 & 13 & 34 [75] & 12 [120] & 17 [290]\\
0 & $1$ & 42 & 680 & 13 & 34 [75] & 12 [120] & 17 [290]\\
\hline\hline
\end{tabular}

\vspace{0.5cm}

\begin{tabular}{|c c || c c| c | c | c|c|}
$\theta^{DL}_{12}$ & $\theta^{DR}_{12}$ & $\Delta m_K$ & $\eps_K $ & $K_L \to \mu^{\pm} e^{\mp}$ & $\mu \to e \gamma$ & $\mu \to eee$ & $ \mu \to e $ in Ti\\
\hline\hline
$\eps$ & 0 & 19 &  310 & 320 & 23 [51] & 11 [110] & 16 [280]\\
$\eps$ & $\eps$ & 3,400 & 54,000 & 320 & 23 [51]  & 14 [140] & 82 [1,400]\\
$\eps$ & $1$ & 4,900 & 80,000 & 970 & 34 [75] & 42 [420]& 250 [4,300]\\
0 & $1$ & 42 & 680 & 970 & 34 [75] & 12 [120] &17 [290]\\
\hline\hline
\end{tabular}

\caption{\label{SU5sector12} 1-2 sector constraints on the messenger scale in TeV for fermion (up) and Higgs (down) UVCs assuming SU(5). 
Values in [ ] give the expected future bounds.}
\end{table}

The results for the 1-2 sector are shown in Table \ref{SU5sector12}. Note that the processes $K_L \to \mu^+ \mu^-$ and $K_L \to e^+ e^-$ do not appear in these tables, since the new physics contribution to these processes is negligible if the constraints on the messenger scale from correlated processes are fulfilled. 

In the case of fermion UVC (Table~\ref{SU5sector12}, up) the strongest bound on the scale comes from $K - \overline{K}$ mixing except for small real angles, in which case $\mu \to e \gamma$ can be more constraining than $\Delta m_K$. The most interesting aspect is that the ratio ${\rm CR}(\mu \to e ~ {\rm in ~Ti})/{\rm BR}(\mu \to e \gamma)$ is always $\gtrsim \ord{0.1}$, much larger than the typical SUSY prediction $\sim\alpha_{em}$. Again this follows from the bounds in Table~\ref{SU5sector12} using
\begin{equation}
\frac{{\rm CR}(\mu \to e~ {\rm in ~Ti})}{{\rm BR}(\mu \to e \gamma)} = \left( \frac{M_{\mu \to e~ {\rm in ~Ti}}}{M_{\mu \to e \gamma}} \right)^4 \,  \frac{{\rm CR^{exp}}(\mu \to e~ {\rm in ~Ti} )}{{\rm BR^{exp}}(\mu \to e \gamma)},
\label{superF2}
\end{equation}
where ${\rm CR^{exp}(\mu \to e~ {\rm in ~Ti} )} = 4.3 \times 10^{-12}$ and the notation is as in Eq.~(\ref{superF}). The correlations in the $\mu-e$ sector are such that, if MEG finds evidence for $\mu \to e\gamma$ with ${\rm BR}(\mu \to e\gamma)\gtrsim 10^{-13}$, then we have using again Eqns.~(\ref{superF}, \ref{superF2})
\begin{align}
{\rm BR}(\mu \to eee) & \gtrsim 8 \times 10^{-16} & {\rm CR}(\mu \to e ~ {\rm in~ Ti}) & \gtrsim 10^{-15},
\end{align}
 i.e.~other LFV processes must be observed at future experiments. If, on the other hand, MEG does not observe $\mu \to e\gamma$ (setting a bound on the scale up to $M\gtrsim 75$ TeV), there is still the possibility to discover $\mu \to eee$ and $\mu \to e$ conversion in Nuclei (since future experiments will test larger scales, up to 120 TeV and 290 TeV respectively). Notice that the future sensitivity for $\mu \to e \gamma$, $\mu \to eee$ and $\mu \to e ~ {\rm in~ Ti}$ is always beyond the bound from $K - \overline{K}$ mixing, provided that the CPV phases are sufficiently suppressed.

In the case of Higgs messengers (Table \ref{SU5sector12}, down), the strongest bound is set either by $K - \overline{K}$ mixing observables or $K_L \to \mu^{\pm} e^{\mp}$, that can now arise at tree-level.  This latter process is the most constraining (even in the CPV case) if at least one of the angles is very small, so that the tree-level contribution to  $K - \overline{K}$ vanishes. The strong bounds from the Kaon sector imply that BR($\mu \to e\gamma$) is always suppressed below the $10^{-16}$ level, i.e.~far beyond the reach of MEG, and $\mu \to eee$ cannot be observed neither in this scenario. The only possible deviations from the SM can be then observed in the Kaon system and for $\mu \to e$ conversion in nuclei, for which can be as large as ${\rm CR}(\mu \to e ~ {\rm in~ Ti}) \simeq 10^{-17}\div 10^{-18}$, provided that CP violating phases are sufficiently suppressed.

Let us briefly comment on the 1-3 and 2-3 sectors. Besides the $B_d$ and $B_s$ mixing observables (respectively $\Delta m_{B_d}$, $S_{\psi K_s}$ and $\Delta m_{B_s}$, $S_{\psi \phi}$) discussed in Table \ref{Bstab}, we consider $B_{d,s}\to \ell^+\ell^-$.\footnote{The LFV $\tau$ decays and $b\to s\gamma$ give negligible bounds on the messenger scale compared to other observables.} For the fermion UVC, all processes in these sectors give a very weak constraint on the messenger scale (below the 1 TeV level), much below the minimal bound of 19 TeV, unless the RH rotations are $\mathcal{O}(1)$. In this case, deviations from the SM are only possible for $\Delta F=2$ observables, since $\Delta m_{B_d}$ and $S_{\psi K_s}$ give at present a bound on the scale of about 25 TeV and the future LHCb sensitivity to $S_{\psi \phi}$ can constrain scales up to 20 TeV (cf.~Table \ref{Bstab}). The same results qualitatively hold in the case of Higgs UVC, but for $B_{d,s}\to \ell^+\ell^-$, that can now arise at tree-level. Larger rates than in the SM are then possible and, in particular, the present experimental bounds for $B_{d,s}\to \mu^+\mu^-$ can be easily saturated. Interestingly, the ratio BR($B_{d}\to \mu^+\mu^-$)/BR($B_{s}\to \mu^+\mu^-$) can be even $\mathcal{O}(1)$, contrary to the SM (and MFV) prediction.

\bigskip
\noindent Let us summarise the phenomenological consequences (valid up to $\ord{1}$ factors) separately for Higgs and fermion UVCs. 
\bigskip

\noindent {\bf \underline{FUVC}}
\begin{itemize}
\item  With the present experimental bounds  the ratio ${\rm CR}(\mu \to e ~ {\rm in ~Ti})/{\rm BR}(\mu \to e \gamma)$ is $\gtrsim \ord{0.1}$.
\item Evidence for $\mu \to e \gamma$ at MEG would imply $\mu \to eee$ and $\mu-e$ conversion in the reach of future experiments.
\item Deviation from the SM in 1-3 and 2-3 transitions can only occur in $\Delta F =2 $ observables.
\end{itemize}


\noindent {\bf \underline{HUVC}}
\begin{itemize}
\item Kaon sector bounds prevent observation of LFV processes except $\mu - e$ conversion.
\item ${\rm BR}(B_{d} \to \mu^+ \mu^-)$ and ${\rm BR}(B_{s} \to \mu^+ \mu^-)$ can saturate present bounds and be comparable to each other, contrary to MFV.
 \end{itemize}


\section{Conclusions}

We have discussed  model-independent  minimal flavour-violating effects induced by  messenger sectors in models of fermion masses and mixing based on horizontal symmetries.
In the flavour basis, integrating out the messengers induces higher-dimensional flavour-violating operators, whose coefficients depend on the specific flavour symmetry and breaking
pattern, and flavour conserving operators. After rotating to the fermion mass basis, the latter also contribute to the flavour-violating operators, providing a minimal contribution to flavour changing neutral currents. This contribution is model independent
in the case of abelian flavour symmetries (and to large extent for the non-abelian case too).
The coefficients of these operators only depend on light rotation angles (i.e.~on the structure of the
Yukawa matrices). In non-abelian models these coefficients are further suppressed by the breaking of flavour universality that is related to small flavon vevs. 

For those minimal, universal contributions to the FCNC and CPV effects we have derived the bounds on the mass scale of messengers.  They are valid (up to $\ord{1}$ coefficients) for any abelian model
and can be easily applied to a large class of non-abelian models taking into account additional suppression factors.
Moreover, the abelian case is relevant for non-abelian models with some SM fermions transforming
as singlets of the non-abelian flavour group. 

The obtained lower bounds on the messenger scale are different for different operators and in addition they depend on the chosen set of rotations.
Given the sensitivity expected in the forthcoming experiments, that leaves interesting room for discovering new physics  and for testing fermion mass models.  As the highlights of our analysis emerge the leptonic processes,  $\mu\rightarrow e \gamma$, $\mu\rightarrow eee$ and $\mu\rightarrow e$ conversion in nuclei.

In more detail, we find that the interplay of $K-\overline{K}$ and $D-\overline{D}$ mixing implies that the messenger scale in abelian models has to be larger
than about 20 TeV. Though quite strong, this minimal bound does not prevent large effects in $B_q - \overline{B}_q$ mixing and in  LFV decays.
In particular, the ratio ${\rm BR}(\mu \to eee )$/${\rm BR}(\mu \to e \gamma)$ can be as large as $\ord{10}$ for large leptonic rotations, contrary to
SUSY scenarios. More generally, the BRs for both processes can be large enough to be within the reach of the future experiments. In non-abelian models the additional suppression factor on 1-2 flavour transitions (at least $\sim \eps^2$) allows messengers at the TeV scale,
and therefore possibly in the reach of LHC.

Assuming approximately SU(5)-symmetric Yukawas, we can include in the analysis relevant $2q2\ell$
operators and provide correlations between quark and lepton sector. We find that for  the UV completion with heavy fermions the ratio ${\rm CR}(\mu \to e ~ {\rm in ~Ti})/{\rm B   R}(\mu \to e \gamma)$ is $\gtrsim \ord{0.1}$.
An evidence for $\mu \to e \gamma$ at MEG would imply $\mu \to eee$ and $\mu-e$ conversion in the reach of future experiments.
Moreover, deviation from the SM in 1-3 and 2-3 transitions can only occur in $\Delta F =2 $ observables.
For the UV completion with heavy Higgses, on the other hand, the bounds from the Kaon sector prevent the observation of LFV processes at running/future
experiments, with the possible exception of $\mu \to e$ conversion in Nuclei.
Also, in this case, the new contributions to $B_{d,s} \to \mu^+ \mu^-$ can saturate the present experimental limits
and give the two processes at comparable rates, contrary to the SM and MFV scenarios.

\section*{Acknowledgements}
We would like to thank S.~Davidson, C.~Hagedorn, G.~Isidori, P.~Paradisi and F.~Sala for useful discussions and communications. 
The Feynman diagrams have been drawn using JaxoDraw \cite{Jaxodraw}.
We thank the Theory Division of CERN for hospitality during several stages of this work. 
L.C.~and R.Z.~are grateful to the Institute of Theoretical Physics of the University of Warsaw for kind hospitality and financial support 
during their stays in Warsaw.
S.P.~and R.Z.~acknowledge support of the TUM-IAS funded by the German Excellence Initiative.
This work has been partially supported by the contract PITN-GA-2009-237920 UNILHC, 
by the National Science Centre in Poland under research grant DEC-2011/01/M/ST2/02466
and by MNiSW under grant N N202 091839.



\begin{thebibliography}{99}

\bibitem{MFV}
  G.~D'Ambrosio, G.~F.~Giudice, G.~Isidori and A.~Strumia,
  Nucl.\ Phys.\ B {\bf 645} (2002) 155
  [hep-ph/0207036].
  
\bibitem{LPR}
  Z.~Lalak, S.~Pokorski, G.~G.~Ross,
  JHEP {\bf 1008}, 129 (2010)
  [arXiv:1006.2375 [hep-ph]].



\bibitem{BGPZ}
  A.~J.~Buras, C.~Grojean, S.~Pokorski and R.~Ziegler,
  JHEP {\bf 1108} (2011) 028
  [arXiv:1105.3725 [hep-ph]].


 \bibitem{High-Energy}
  L.~Calibbi, Z.~Lalak, S.~Pokorski and R.~Ziegler,
  arXiv:1203.1489 [hep-ph].

\bibitem{Grinstein}
  B.~Grinstein, M.~Redi and G.~Villadoro,
  JHEP {\bf 1011} (2010) 067
  [arXiv:1009.2049 [hep-ph]].

\bibitem{Andrzej}
  A.~J.~Buras, M.~V.~Carlucci, L.~Merlo and E.~Stamou,
  arXiv:1112.4477 [hep-ph].
    
\bibitem{FN}
  C.~D.~Froggatt and H.~B.~Nielsen,
  Nucl.\ Phys.\  B {\bf 147} (1979) 277.

\bibitem{MMM1}
M.~Leurer, Y.~Nir, N.~Seiberg,
  Nucl.\ Phys.\  {\bf B398 } (1993)  319-342
  [hep-ph/9212278].
  
  \bibitem{MMM2}
 M.~Leurer, Y.~Nir, N.~Seiberg,
  Nucl.\ Phys.\  {\bf B420}, 468-504 (1994)
  [hep-ph/9310320].

\bibitem{Berezhiani:1990wn}
  Z.~G.~Berezhiani and M.~Y.~.Khlopov,
  Sov.\ J.\ Nucl.\ Phys.\  {\bf 51} (1990) 739
   [Yad.\ Fiz.\  {\bf 51} (1990) 1157].



\bibitem{King-Ross}
  S.~F.~King and G.~G.~Ross,
  Phys.\ Lett.\  B {\bf 520} (2001) 243
  [arXiv:hep-ph/0108112].


\bibitem{u2-models}
  A.~Pomarol and D.~Tommasini,
  Nucl.\ Phys.\ B {\bf 466} (1996) 3
  [hep-ph/9507462];
  R.~Barbieri, G.~R.~Dvali and L.~J.~Hall,
  Phys.\ Lett.\ B {\bf 377} (1996) 76
  [hep-ph/9512388].

\bibitem{AF} 
  G.~Altarelli and F.~Feruglio,
  Rev.\ Mod.\ Phys.\  {\bf 82}, 2701 (2010)
  [arXiv:1002.0211 [hep-ph]].
  
\bibitem{Feruglio} 
  F.~Feruglio, C.~Hagedorn, Y.~Lin and L.~Merlo,
  Nucl.\ Phys.\ B {\bf 809} (2009) 218
  [arXiv:0807.3160 [hep-ph]];
   F.~Feruglio and A.~Paris,
  Nucl.\ Phys.\ B {\bf 840}, 405 (2010)
  [arXiv:1005.5526 [hep-ph]].

\bibitem{GinoNir}
  G.~Isidori, Y.~Nir and G.~Perez,
  Ann.\ Rev.\ Nucl.\ Part.\ Sci.\  {\bf 60} (2010) 355
  [arXiv:1002.0900 [hep-ph]].

\bibitem{PerezTASI}
  O.~Gedalia and G.~Perez,
  arXiv:1005.3106 [hep-ph].

\bibitem{PDG}
  K.~Nakamura {\it et al.}  [Particle Data Group Collaboration],
  J.\ Phys.\ G G {\bf 37} (2010) 075021.

\bibitem{DDb}
  M.~Bona {\it et al.}  [UTfit Collaboration],
  JHEP {\bf 0803} (2008) 049
  [arXiv:0707.0636 [hep-ph]];
  G.~Isidori, J.~F.~Kamenik, Z.~Ligeti and G.~Perez,
  arXiv:1111.4987 [hep-ph].


\bibitem{LHCbpsiphi} 
R.~Aaij {\it et al.}  [LHCb Collaboration],
  Phys.\ Rev.\ Lett.\  {\bf 108}, 101803 (2012)
  [arXiv:1112.3183 [hep-ex]].
 
\bibitem{HFAG}
  D.~Asner {\it et al.}  [Heavy Flavor Averaging Group Collaboration],
  arXiv:1010.1589 [hep-ex], and online update at http://www.slac.stanford.edu/xorg/hfag.


\bibitem{LHCbupgrade}
LHCb Collaboration,
``Letter of Intent for the LHCb Upgrade,"  
CERN-LHCC-2011-001; LHCC-I-018.

\bibitem{MEG}
  J.~Adam {\it et al.}  [MEG Collaboration],
  Phys.\ Rev.\ Lett.\  {\bf 107} (2011) 171801
  [arXiv:1107.5547 [hep-ex]].

\bibitem{PSI}
  A.~Sch\"oning, S.~Bachmann and R.~Narayan,
  Physics Procedia {\bf 17} (2011) 181.
  

\bibitem{BurasParide}
  W.~Altmannshofer, A.~J.~Buras, S.~Gori, P.~Paradisi and D.~M.~Straub,
  Nucl.\ Phys.\ B {\bf 830} (2010) 17
  [arXiv:0909.1333 [hep-ph]].

\bibitem{GJ}
  H.~Georgi and C.~Jarlskog,
  Phys.\ Lett.\ B {\bf 86} (1979) 297.

\bibitem{Davidson}
  M.~Carpentier and S.~Davidson,
  Eur.\ Phys.\ J.\ C {\bf 70} (2010) 1071
  [arXiv:1008.0280 [hep-ph]].

\bibitem{LHCb}
  R.~Aaij {\it et al.}  [LHCb Collaboration],
  arXiv:1203.4493 [hep-ex].


\bibitem{mec}
Talk by T.~Mori at EPS-HEP 2011 conference, Grenoble.


\bibitem{Jaxodraw}
  D.~Binosi, J.~Collins, C.~Kaufhold and L.~Theussl,
  Comput.\ Phys.\ Commun.\  {\bf 180} (2009) 1709
  [arXiv:0811.4113 [hep-ph]];
  D.~Binosi and L.~Theussl,
  Comput.\ Phys.\ Commun.\  {\bf 161} (2004) 76
  [hep-ph/0309015].



\end{thebibliography}
\end{document}